\documentclass[letters,usenatbib]{mnras}

\usepackage{mathptmx}

\usepackage[T1]{fontenc}
\usepackage{ae,aecompl}

\usepackage{graphicx}	
\usepackage{amsmath}	
\usepackage{amssymb}	
\usepackage{multirow}

\usepackage{times}

\usepackage[usenames,dvipsnames]{xcolor}

\newcommand{\lowcase}{\scshape}

\title[New $\gamma$-ray-emitting NLS1 at Redshift $\sim1$]{
Identification of a New $\gamma$-ray-emitting narrow-line Seyfert 1 Galaxy, at Redshift $\sim1$
}

\author[S. Yao et al.]{
Su Yao,$^{1,2,3}$\thanks{E-mail: yaosu@nao.cas.cn}
Weimin Yuan,$^{1}$\thanks{E-mail: wmy@nao.cas.cn}
Hongyan Zhou,$^{2,4}$
S. Komossa,$^{5}$
Jin Zhang,$^{1}$
\newauthor Erlin Qiao$^{1}$
and 
Bifang Liu$^{1}$
\\
$^{1}$Key Laboratory of Space Astronomy and Technology, National Astronomical Observatories, Chinese Academy of Sciences, \\Beijing 100012, China\\
$^{2}$Polar Research Institute of China, 451 Jinqiao Road, Pudong, Shanghai 200136, China\\
$^{3}$University of Chinese Academy of Sciences, Beijing 100049, China\\
$^{4}$Key Laboratory for Research in Galaxies and Cosmology, Department of Astronomy, University of Science and Technology of China, \\Chinese Academy of Science, Hefei, Anhui 230026, China\\
$^{5}$Max-Planck Institut f{\"u}r Radioastronomie, Auf dem H{\"u}gel 69, 
53121 Bonn, Germany\\
}

\date{Accepted XXX. Received YYY; in original form ZZZ}

\pubyear{2015}

\begin{document}
\label{firstpage}
\pagerange{\pageref{firstpage}--\pageref{lastpage}}
\maketitle

\begin{abstract}

We report on the identification of a new $\gamma$-ray-emitting narrow-line Seyfert 1 (NLS1) galaxy, SDSS~J122222.55+041315.7, which increases the number of 
known objects of this remarkable but rare type of active galactic nuclei (AGN) to seven.
Its  optical spectrum, obtained in the Sloan Digital Sky Survey-Baryon Oscillation Spectroscopic Survey, 
reveals a broad H\,$\beta$ emission line with a width (FWHM) 
of 1734$\pm$104\,km\,s$^{-1}$. 
This, along with strong optical Fe{\scshape\,ii} multiplets [$R_{4570}=0.9$] 
and a weak [O{\scshape~iii}]~$\lambda 5007$ emission line, 
makes the object a typical NLS1.  
On the other hand, the source exhibits a high radio brightness temperature, 
rapid infrared variability, and a flat X-ray spectrum extending up to $\sim$200\,keV. 
It is  associated with a luminous $\gamma$-ray source detected 
significantly with {\it Fermi}/LAT. 
Correlated variability with other wavebands has not yet
been tested. 
The spectral energy distribution can be well modelled by a one-zone leptonic jet model. 
This new member is by far the most distant $\gamma$-ray-emitting  NLS1, at a redshift of $z=0.966$.

\end{abstract}



\section{Introduction}

With the full width at half-maximum (FWHM) of the broad H\,$\beta$ line less than 2000\,km\,s$^{-1}$, narrow-line Seyfert 1 galaxies (NLS1s\footnote{The term `NLS1s' here stands collectively for both high-luminosity narrow-line type I quasars and low-luminosity NLS1s. }) have the strongest Fe{\,\lowcase ii} emission and the weakest [O{\,\scshape iii}] emission among active galactic nuclei (AGNs). 
They occupy an extreme end of the so-called `Eigenvector-I' (EV1) parameter space, a set of correlations between the H\,$\beta$ line width and other observables \citep[e.g.][]{BG92}. 
It has been shown in previous work that 
radio loud (RL) NLS1s are rare, 
only $\sim$7\% in fraction \citep[e.g.][]{2006AJ....132..531K, ZhouHY2006}, 
compared to about 10--15 per cent in broad-line Seyfert 1 galaxies and quasars \citep[e.g.][]{2002AJ....124.2364I}. 
The very RL NLS1s with radio loudness $R>100$ are even rarer, accounting for only 2.5\% \citep[][]{2006AJ....132..531K}. 
The reason for the rarity of RL NLS1s is still unclear.

In contrast to NLS1s, the bulk of RL AGNs lies at the other extreme of the EV1 parameter space, 
having broader H\,$\beta$ and weaker Fe{\,\lowcase ii} emission \citep[e.g.][]{Sulentic2008}. 
As a distinct subclass, blazars are believed to be RL AGNs viewed with their relativistic jets closely aligned to the line of sight. 
They are characterized by flat radio spectra, compact radio cores, high brightness temperatures and rapid variability in multi-wave bands \citep[e.g.][]{1995PASP..107..803U}.

Since NLS1s and blazars lie at the opposite extremes in the AGN parameter space, 
the discovery of their hybrids 
in the past decade \citep[e.g.,][]{2003ApJ...584..147Z, 2007ApJ...658L..13Z}
has drawn considerable attention. 
By performing a comprehensive study of 23 very RL NLS1s, \citet[][]{2008ApJ...685..801Y} 
pointed out the peculiarity of these objects and 
suggested that there exists a population of RL NLS1s possessing relativistic jets, similar to blazars.  
This assertion was confirmed and highlighted by the detection of several objects 
in the GeV $\gamma$-rays with {\it Fermi}/LAT 
\citep[e.g.,][]{2009ApJ...699..976A, 2009ApJ...707L.142A, 2011nlsg.confE..24F, 2012MNRAS.426..317D}. 
These discoveries raised questions regarding the coupling of jet and accretion flow \citep[][]{2015AJ....150...23Y}. 
However, further investigation of this class of objects in large numbers is hampered by their scarcity. 
There are only six NLS1s detected in the $\gamma$-ray band by {\it Fermi}/LAT at high significance hitherto\footnote{Two more tentative candidates, FBQS~J1102+2239 and SDSS~J1246+0238, were briefly reported by \citet{2011nlsg.confE..24F}. } 
(PMN~J0948$+$0022, \citealt{2009ApJ...699..976A}; PKS~1502$+$036, 1H~0323$+$342 and PKS~2004$-$447, \citealt{2009ApJ...707L.142A}; SBS~0846$+$513, \citealt{2012MNRAS.426..317D}; FBQS~J1644+2619, \citealt{2015MNRAS.452..520D}), 
mostly from the \citet[][]{2008ApJ...685..801Y} sample.

In this Letter, we report the identification of a {\it Fermi}/LAT detected flat-spectrum radio quasar (FSRQ) 
with a new $\gamma$-ray-emitting NLS1, SDSS~J122222.55+041315.7 
($z=0.966$, hereafter J1222+0413). 
This object was found to be an NLS1, and the only NLS1 associated with a {\it Fermi}/LAT $\gamma$-ray source, in the course of our ongoing program to search for RL NLS1s from the Sloan Digital Sky Survey-Baryon Oscillation Spectroscopic Survey \citep[SDSS-BOSS;][]{2014ApJS..211...17A}. 
J1222+0413 was classified as an FSRQ in previous surveys 
mainly based on its broad Mg{\scshape~ii} emission line 
\citep[e.g.][]{2012RMxAA..48....9T}. 
However, its NLS1 nature was not known due to the lack of a spectrum covering the H\,$\beta$ line given its relatively high redshift. 
Here we present an analysis of its optical spectrum obtained from 
SDSS-BOSS
whose bandpass extending to $\sim$10000\,\AA~covers the redshifted H\,$\beta$ region. 
This finding increases the number of the very rare $\gamma$-ray-emitting NLS1s to seven. 
Throughout this work a cosmology is assumed with $H_0=70$ km s$^{-1}$ Mpc$^{-1}$, $\Omega_\Lambda=0.73$ and $\Omega_{\rm M}=0.27$.

\section{Optical Spectroscopy}
\label{spectroscopy}

\begin{figure}
	\includegraphics[width=\columnwidth]{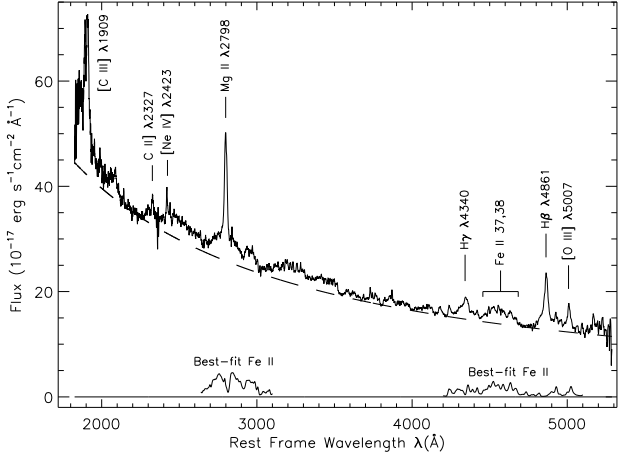}
	\caption{Rest-frame BOSS-SDSS spectrum of J1222+0413 after correction for the Galactic extinction. 
	The dashed line represents a single power-law continuum. 
	}
	\label{sdss_spec}
\end{figure}
The spectrum (Fig.~\ref{sdss_spec}) is first corrected for Galactic extinction with $E(B-V)$=0.016 mag \citep{2011ApJ...737..103S} and an $R_V$=3.1 extinction law, and then are transformed into the source rest frame at a redshift of $z$=0.966 provided by the SDSS spectroscopic pipeline. 
The host galaxy contribution is negligible.

We adopt a similar approach as in \citet[][]{2008MNRAS.383..581D} to fit simultaneously the continuum, the Fe{\scshape~ii} multiplets and other emission lines of the spectrum in the ranges of 2690--3100\,\AA~and 4200--5100\,\AA. 
A single power law is used to model the sum of the thermal continuum emission and the contribution from non-thermal emission of the jet.
The optical and ultraviolet Fe{\scshape~ii} are modelled using the same templates as in \citet{2008MNRAS.383..581D} and \citet{WangJG2009}, respectively. 
To fit the Balmer lines, 
H\,$\beta$ and H$\gamma$ are assumed to have the same redshift and profile. 
The broad component of each is accounted for by either a single Lorentzian or concentric double-Gaussian. 
In addition, another Gaussian with constraint of FWHM$<$900\,km\,s$^{-1}$ is used for modelling the narrow component of the Balmer lines. 
In a first step, 
the [O\,{\scshape iii}]\,$\lambda\lambda$4959,\,5007 doublet is modelled with a single Gaussian for each line. 
The flux ratio of [O\,{\scshape iii}]\,$\lambda\lambda4959,5007$ is fixed at the theoretical value of 1:3. 
Each of the two lines of the Mg{\scshape\,ii}\,$\lambda\lambda$2796,\,2803 doublet is modelled with two components, one Lorentzian for the broad component and one Gaussian for the narrow. 
The broad components of the Mg{\scshape\,ii} doublet are set to have the same profile, with their flux ratio fixed at 1.2:1 assuming an entirely thermalized gas for simplicity \citep[e.g.][]{1997ApJ...489..656L}. 
The same prescription is applied to the narrow components of Mg{\scshape\,ii} with an additional constraint of FWHM$<$900\,km\,s$^{-1}$.

Then the models of the continuum, the Fe{\scshape~ii} and the emission lines are fitted simultaneously to the spectrum. 
We find that the total H\,$\beta$ line is similarly well fitted with either Lorentzian$+$Gaussian profile or double-Gaussian$+$Gaussian profile. 
In the former case, the width of the broad component is FWHM(H$\beta_{\rm broad}$)$=1734\pm$104\,km\,s$^{-1}$ while in the latter case FWHM(H$\beta_{\rm broad}$)$=2264\pm$350\,km\,s$^{-1}$. 
A direct measurement of the total H\,$\beta$ line gives FWHM(H$\beta_{\rm total}$)$=1576$\,km\,s$^{-1}$. 
These widths put J1222+0413 within the NLS1 regime, near the border line between NLS1 and normal broad-line Seyfert 1 (BLS1). 
In the following we use the Lorentzian representation of the broad component, 
because it has been shown that the broad H\,$\beta$ line of NLS1s is generally better fitted with a Lorentzian profile \citep[e.g.,][]{2001A&A...372..730V, ZhouHY2006}. 
In this case, the narrow component of H\,$\beta$ is relatively faint, with FWHM$\approx$800\,km\,s$^{-1}$. 
The line ratio of [O{\,\scshape iii}]\,$\lambda5007$ to H$\beta_{\rm total}$ is $\approx0.2$. 
The Fe{\scshape\,ii} is strong, $R_{4570}\equiv$Fe\,{\scshape ii}\,$\lambda4570$/H$\beta_{\rm total}\approx0.9$ where Fe{\scshape\,ii}\,$\lambda4570$ is calculated by integrating from 4434 to 4684\,\AA. 
These features clearly fulfill the conventional definition of an NLS1. 

\begin{figure}
	\centering
	\includegraphics[width=\columnwidth]{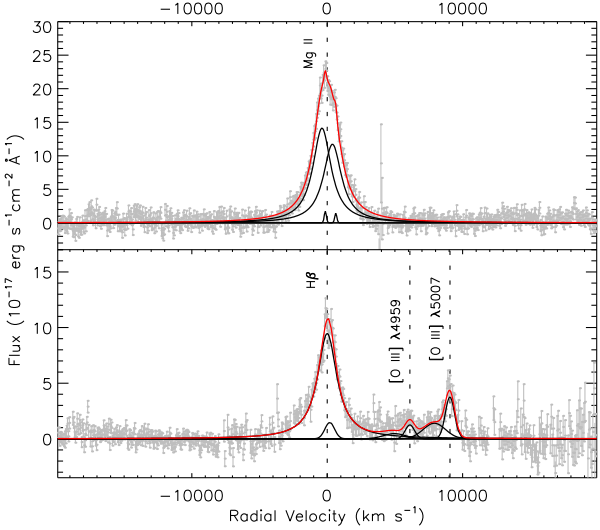}
	\caption{
	Residuals (in grey) of the spectrum in the Mg{\scshape\,ii} region ({upper panel}) and H$\beta+$[O{\scshape\,iii}] region ({lower panel}) after subtracting the power-law continuum and the Fe{\scshape\,ii} model. 
	The best-fitting Lorentzians$+$Gaussians are displayed in black; their sum is in red. 
	\label{sdss_lines}}
\end{figure}

Fig.~\ref{sdss_lines} shows the residuals of the spectrum after subtracting the power-law continuum and the Fe{\scshape\,ii} model. 
We notice that an asymmetric blue wing of H\,$\beta$ is present. 
We also notice that [O{\scshape\,iii}]\,$\lambda5007$ is relatively broad when fitted by a single Gaussian (FWHM$\approx$1400\,km\,s$^{-1}$). 
This may be the effect of an outflow in J1222+0413. 
It was shown in previous studies that 
NLS1s often show strong blue wings and blueshifts in [O{\scshape\,iii}] \citep[e.g.][]{2007ApJ...667L..33K}. 
Thus, in a second step, we used two Gaussians to model each line of the [O{\scshape\,iii}] doublet, one for a core and the other for a blueshifted wing component, 
while the continuum and other emission lines are fixed at the best-fitted results as above. 
Although the [O{\scshape\,iii}] spectrum is located near the red end of the spectrum and has low signal-to-noise ratio (S/N) due to the telluric absorption, 
we find a broad blue wing, characterized by FWHM$\sim$1800\,km\,s$^{-1}$, which is blueshifted by 1100\,km\,s$^{-1}$ from the narrow core component. 
The fitting results of major emission lines are summarized in Table~\ref{best_params}. 

\begin{table}
	\caption{Parameters of the main optical emission lines. }
	\label{best_params}
	\begin{center}
	\begin{tabular}{lccc}
		\hline
		 Line & Model$^a$ & FWHM$^b$ & Flux  \\
		         &           & (km s$^{-1}$) & (erg s$^{-1}$cm$^{-2}$)  \\
		\hline
		H$\beta_{\rm broad}$ & L & $1734\pm104$ & $8.1\times10^{-15}$ \\
		H$\beta_{\rm broad}$ & 2G & $2264\pm350$ & $6.1\times10^{-15}$ \\
		H$\beta_{\rm total}$   & D & 1530 & $8.6\times10^{-15}$ \\
		$[$O~{\scshape iii}]$\lambda5007_{\rm core}$ & G & $791\pm125$ & $1.0\times10^{-15}$ \\
		$[$O~{\scshape iii}]$\lambda5007_{\rm wing}$ & G & $1841\pm671$ & $9.0\times10^{-16}$ \\
		Mg{\scshape\,ii}$_{\rm broad}$ & L & $1638\pm45$ & $12.1\times10^{-15}$ \\
		\hline
	\end{tabular}
	\parbox[]{\columnwidth}{
	$^a$Models used for fitting the emission lines are: one Lorentzian profile (L), one Gaussian profile (G) or a Double-Gaussian profile (2G). `D' stands for the direct integration over the observed line profile. \\
	$^b$FWHMs are not corrected for instrumental resolution. 
	}
	\end{center}
\end{table}


\section{BLAZAR-LIKE PROPERTIES}

\subsection{Radio emission}

J1222+0413 was detected in several radio surveys.  
The spectral index
between 1.4 and 4.85\,GHz is flat, $\alpha_{\rm rad}=0.3$ \citep[$S_{\nu}\propto\nu^{\alpha}$,][]{1992ApJS...79..331W}. 
The core flux density at 1.4\,GHz is 0.6\,Jy \citep{2010ApJ...710..764K}, corresponding to a radio power of $P_{\rm 1.4GHz}\sim1\times10^{27}$\,W\,Hz$^{-1}$. 
\citet{2012A&A...544A..34P} investigated the 
very long baseline interferometry 
images at 2.3 and 8.6\,GHz, and estimated a core brightness temperature of $1.43\times10^{12}$ and $4.34\times10^{12}$\,K, respectively, clearly indicating Doppler-boosted emission. 
As a blazar, J1222+0413 is expected to be variable in the radio band. 
We have checked the radio data in NASA/IPAC Extragalactic Database (NED) and literatures. 
The total flux densities varied in a range of $\sim$0.66--0.8\,Jy at 1.4\,GHz and in a range of $\sim$0.49--1.08\,Jy at 5\,GHz on times-cales from years to decade. 
The most intensive radio monitoring of J1222+0413 was performed at 15\,GHz with cadence as short as two days, 
yielding the maximum variability amplitude of 24\% within 405\,d \citep[206\,d in the object's rest frame, see][]{2011ApJS..194...29R}. 

To estimate the radio loudness $R_{5\rm GHz}\equiv f_{\nu}(5{\rm GHz})/f_{\nu}$(4400{\AA}), 
we use the 5\,GHz core flux \citep[665\,mJy,][]{1997A&AS..122..235L} and the SDSS $g$-magnitude, by considering the $k$-correction with radio spectral index $\alpha_{\rm rad}=0.3$ and the optical spectral index $\alpha_{\rm opt}=-0.72$ obtained from the spectral fitting (Section~\ref{spectroscopy}). 
The result is $R_{5\rm GHz}\sim1700$. 
Even considering the radio variability, J1222+0413 is still a very RL AGN.

\subsection{Rapid infrared variability}

J1222+0413 was observed with the {\it Wide-field Infrared Survey Explorer} ({\it WISE}) 
in four bands $w1$, $w2$, $w3$ and $w4$ centred at 3.4, 4.6, 12 and 24\,$\mu$m, respectively. 
The variability flag in the {\it WISE} All-Sky Source Catalog {\it var\_flag} of ``9910'' indicates that this source has a very high probability of being variable in the $w1$ and $w2$ bands \citep[e.g.,][]{2012ApJ...759L..31J}. 

\begin{figure}
	\centering
	\includegraphics[width=\columnwidth]{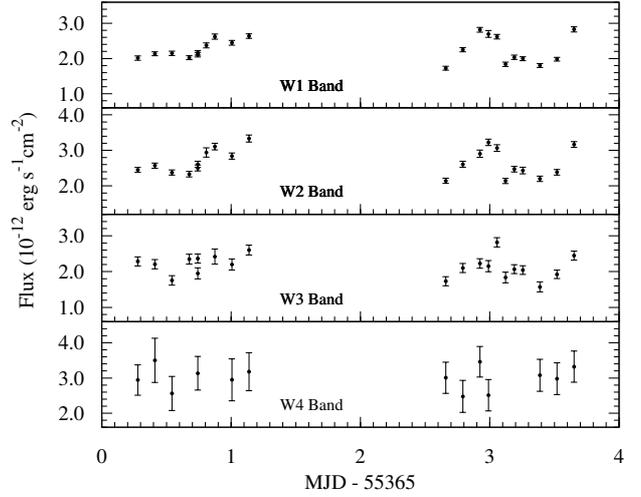}
	\caption{{\it WISE} light curves in the $w1$, $w2$, $w3$ and $w4$ band ({from top to bottom panel}) constructed from profile-fit magnitudes with 1\,$\sigma$ error. }
	\label{wise_lc}
\end{figure}

Its light curves are constructed from the PSF profile-fit photometric magnitudes and converted to fluxes \citep[][]{2010AJ....140.1868W} after excluding the data whose S/N is marked as ``null''. 
We also exclude those data with S/N$<$10 for the $w1$ and $w2$ bands, and S/N$<$5 for the $w3$ and $w4$ bands, and those with reduced $\chi^2$ of the profile-fit photometries larger than 2. 
The light curves are shown in Fig.~\ref{wise_lc}. 
Variability in the $w1$, $w2$ and $w3$ light curves are significantly detected within one day, with $p$-values of $P<0.1\%$ using the $\chi^2$-test against the null hypothesis of no variation. 
This time-scale restricts the size of the emitting region to $\lesssim8\times10^{-4}$\,pc, much smaller than the scale of the torus but consistent with the size of the jet emitting region. 
The variability of $w4$ is not detected possibly due to the lower S/N and much larger photometric errors than the other three bands.

\subsection{High-energy emission} 
\label{xray}

Using the {\it XMM-Newton} and {\it INTEGRAL} data,  
\citet{2008MNRAS.388L..54D} found a hard power-law continuum with a photon index of $\Gamma\sim1.2$ for the object's X-ray spectrum in the broad 2--150\,keV band. 
A soft X-ray excess was also found below 2\,keV which can be described by a blackbody with $kT\sim0.15$\,keV.

{\it Swift} performed six observations on J1222+0413 from 2007 August to 2011 June. 
We reduced the data using {\it {\it Swift}} {\sc ftools} in {\sc heasoft\,v6.15} following the standard procedures \citep[e.g.,][]{2015AJ....150...23Y}. 
Due to relatively poor statistics, the x-ray spectra are consistent with a single power law with absorption fixed at the Galactic value 
\citep[$N^{\rm Gal}_{\rm H}=1.63\times10^{20}\,$cm$^{-2}$,][]{2005A&A...440..775K}. 
The photon indices, $\Gamma=$1.1--1.5, are too flat for normal radio-quiet NLS1s ($\Gamma\sim$2-4, e.g., \citealt{2010ApJS..187...64G, 2011ApJ...727...31A}), but very similar to the other $\gamma$-ray-emitting NLS1s \citep[$\Gamma<2$, e.g.,][]{2009ApJ...707L.142A}. 
The 0.3--10\,keV fluxes varied by a factor of 2 during the observations, but the correlation between the observed indices and the fluxes are not detected. 
This object is also detected with the {\it Swift}/BAT in the 14--195\,keV band, as given in the {\it Swift}/BAT 70\,Month Catalog \citep[][]{2013ApJS..207...19B}. 
The average hard X-ray spectrum is also flat, with a photon index of $\Gamma=$1.3. 
Compared to the previous {\it INTEGRAL} observations taken several years ago \citep{2008MNRAS.388L..54D}, 
the hard X-ray data show almost no significant variability 
(see Fig.~\ref{sed}). 
Such a flat spectrum extending to the hard X-ray band is characteristic of relativistic jet emission. 

In the $\gamma$-ray band, J1222+0413 was reported to be associated with the $\gamma$-ray source 3FGL\,J1222.4+0414 in the Third {\it Fermi}/LAT AGN Catalog (3LAC), 
with $\gamma$-ray detection significance of 30\,$\sigma$ \citep[][]{2015ApJ...810...14A}. 
The flux in the band 100\,MeV--100\,GeV is $2.9\times10^{-11}$\,erg\,cm$^{-2}$\,s$^{-1}$, corresponding to an isotropic $\gamma$-ray luminosity of $1.4\times10^{47}$\,erg\,s$^{-1}$. 
The photon index of the $\gamma$-ray spectrum is 2.77 when assuming a single power-law slope, similar to those of the other $\gamma$-ray-emitting NLS1s which are in the range of 2.2--2.8 \citep[e.g.,][]{2009ApJ...707L.142A, 2012MNRAS.426..317D}, as well as those of FSRQs which are in a range of 2--3 \citep[e.g.,][]{2015ApJ...810...14A}. 
Note that $\gamma$-ray sources typically are considered as 
confirmed counterparts only, if correlated variability in other wavebands is detected, which is not yet seen in J1222+0413. 
However, we note that the hard X-ray spectrum of J1222+0413 is very flat, and its extension to the higher energies therefore is consistent with its $\gamma$-ray detection.

\subsection{Broad Band SED}
\label{section_sed}

The  broad-band spectral energy distribution (SED) of J1222+0413, 
based on non-simultaneous data up to only 150\,keV (observed with {\it INTEGRAL}),
was  modelled by \citet{2008MNRAS.388L..54D} using  a simple one-zone jet model.
\citet{2012A&A...541A.160G} used simultaneous {\it Planck}, {\it Swift} and {\it Fermi} observations to build the SED of J1222+0413 and fitted it with a third-degree polynomial.
These SEDs show two broad humps, characteristic of the jet emission from blazars.

\begin{figure}
	\centering
	\includegraphics[width=\columnwidth]{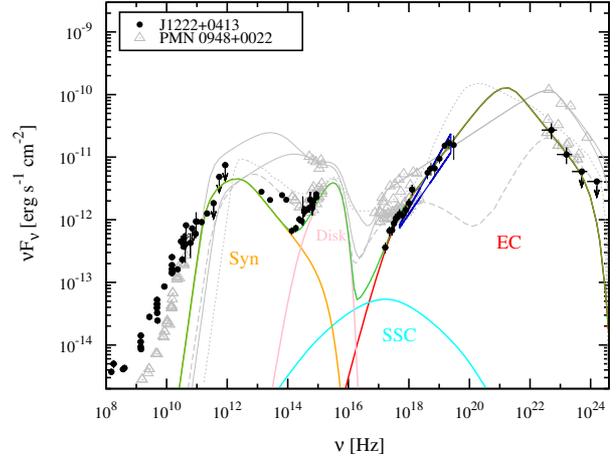}
	\caption{
	The SED of J1222+0412 and its model. 
	The green line represents the sum of all contributions from the synchrotron (syn), synchrotron-self Compton (SSC) and External Compton (EC) processes, and disc component. 
	The data are taken with {\it WISE}, SDSS, {\it Swift}/UVOT, XRT, BAT and {\it Fermi}/LAT \citep{2012A&A...541A.160G}. 
	The blue trapezoid represents the X-ray spectrum of {\it XMM-Newton} and {\it INTEGRAL} in \citet{2008MNRAS.388L..54D}. 
	The SED of another $\gamma$-ray-emitting NLS1, PMN~J0948+0022, and its models at different epochs taken from \citet[][grey solid lines]{2015MNRAS.446.2456D}, \citet[][grey dashed line]{2015ApJ...798...43S} and \citet[][grey dotted lines]{2012A&A...548A.106F} are overploted for comparison. 
	\label{sed}}
\end{figure}

Here we show the SED of J1222+0413 compiled using non-simultaneous data (see Fig.~\ref{sed} and its caption). 
We fit it with a 
one-zone leptonic jet plus a standard disc model\footnote{It should be noted that we adopt the standard disc to model the accretion flow for simplicity, which is an approximation in cases of high Eddington ratios. } 
using the same method as described in \citet[][]{2015ApJ...807...51Z}. 
We consider two cases for the  external Compton (EC) scattering process respectively,
a seed photon field dominated by that from the torus which is expected to be present in any AGN, and by that from the broad line region (BLR).
The energy density of the former is assumed to be $10^{-4}$\,erg\,cm$^{-3}$ \citep[][]{2007ApJ...660..117C} and the latter is estimated from the broad H\,$\beta$ and Mg{\sc~ii} line luminosities derived in Section~\ref{spectroscopy} (see \citealt[][]{2015ApJ...807...51Z} for details). 
We find that the EC/torus model gives a better fit to the high-energy data 
compared to the EC/BLR, and is thus considered to be the more likely model here,
though the latter cannot be ruled out given the non-simultaneity of the data. 
The fitted bulk Lorentz factor of the jet is $\Gamma_{\rm jet}=35$ for the EC/torus case and $\Gamma_{\rm jet}=26$ for the EC/BLR case.

The best-fitting SED model (EC/torus) is shown  in Fig.~\ref{sed}. 
As can be seen, 
the optical-ultraviolet band is dominated by thermal emission from the accretion disc, 
whereas the radio-to-infrared and the X-ray to $\gamma$-ray bands are dominated by the jet
synchrotron 
and EC  radiation, respectively.

\section{DISCUSSION}
\label{discussion}

By analysing the BOSS spectrum of J1222+0413, 
we have shown that the optical spectrum of this blazar is typical of an NLS1, very similar to PMN~J0948+0022 -- the prototype of this class \citep[][]{2003ApJ...584..147Z}. 
Our finding increases the number of known NLS1s with highly significant $\gamma$-ray detection to seven. 
With $z$=0.966, it is the highest-redshift object of this kind detected so far. 
To estimate the black hole mass of J1222+0413, we adopt FWHM(H$\beta_{\rm broad}$)$=1734$\,km\,s$^{-1}$ and the H\,$\beta$-based formalism given in \citet[][]{WangJG2009}, by assuming that the BLR is virial as in \citet[][]{2008ApJ...685..801Y}. 
The 5100\,\AA~luminosity $\lambda L_{5100}$ is estimated from the broad H\,$\beta$ line luminosity using equation 5 of \citet{ZhouHY2006} to eliminate the effect of the jet contribution. 
We find $M_{\rm BH}\approx1.8\times10^8\,M_{\odot}$ ($M_{\rm BH}\approx2.0\times10^8\,M_{\odot}$ if the line parameters fitted from the double Gauissian profile are used), 
consistent with the estimation using Mg{\scshape\,ii} by \citet[][]{2012MNRAS.421.1764S}. 
Similar $M_{\rm BH}$ values are found when using other commonly used one-epoch formalisms, such as in \citet[][]{2006ApJ...641..689V}. 
We note that this value is higher than the typical masses for NLS1s \citep[$10^{6-7}M_{\odot}$;][]{ZhouHY2006, 2012AJ....143...83X}, and in fact falls between the bulk distributions of NLS1s and blazars \citep[e.g.][]{2005ApJ...631..762W}. 
It is also higher than most of the $\gamma$-ray detected NLS1s, but similar to that of PMN~J0948+0022 \citep[$\sim10^8M_{\odot}$;][]{2003ApJ...584..147Z}.

We estimate the bolometric luminosity by assuming $L_{\rm bol}=9\lambda L_{5100}$ (estimated from H\,$\beta$), as suggested by \citet{2000ApJ...533..631K}, and find $L_{\rm bol}\approx1.4\times10^{46}$\,erg\,s$^{-1}$. 
The Eddington ratio of this object is $\lambda\approx0.6$, also typical of NLS1s
which usually have systematically high $\lambda$ close to 1 
\citep[see][and references therein]{2008RMxAC..32...86K}. 
It is worth noting that almost all the $\gamma$-ray detected NLS1s have black holes 
accreting at near the Eddington rate \citep[][]{2009ApJ...707L.142A}. 
They may be analougs to the radio-bright `very high state' in black hole X-ray binaries (e.g., \citealt{2004MNRAS.355.1105F}), during which episodic jets are formed
(see discussion in \citealt{2008ApJ...685..801Y}, and also \citealt[][]{2015A&A...575A..13F}). 

Given the close similarities in the emission-line spectra and the black hole mass
between J1222+0413 and  PMN~J0948+0022, 
it would be interesting to compare their SEDs.  
The averaged SED of PMN~J0948+0022
and its model fits to the data taken at various epochs as given in \citet[][]{2015MNRAS.446.2456D}, \citet{2015ApJ...798...43S} and \citet[][]{2012A&A...548A.106F}
are over-plotted in Fig.~\ref{sed}  for comparison. 
As can be  seen, 
the  X-ray to $\gamma$-ray emission of J1222+0413  forms a single peak which can be well
modelled by a strong EC component (this holds for both the EC/torus and EC/BLR cases). 
The SEDs of PMN~J0948+0022 at different epochs can be equally well modelled by either a single or double peak, due to either lack of the hard X-ray detection for PMN~J0948+0022 and the adoption of different models of EC/torus and EC/BLR cases (see \citealt[][]{2015MNRAS.446.2456D} and \citealt[][]{2015ApJ...798...43S} for details). 
The bulk Lorentz factor of the jet for J1222+0413, 
$\Gamma_{\rm jet}=35$, 
according to our best-fitting model
(or $\Gamma_{\rm jet}=26$ in the EC/BLR case) 
is similar to the value derived by \citet[][$\Gamma_{\rm jet}=30$]{2015MNRAS.446.2456D} in the EC/torus case for PMN~J0948+0022, slightly higher than those in other $\gamma$-ray detected NLS1s \citep[$\Gamma_{\rm jet}\le15$, e.g.,][]{2009ApJ...707L.142A, 2012MNRAS.426..317D}. 
Discoveries of more $\gamma$-ray-emitting NLS1s, 
especially at higher redshifts, 
will provide us with new constraints on the physics of the formation of relativistic jets in this intriguing class of objects 
and their cosmic evolution.

\section*{Acknowledgements}
We thank Tuo Ji, 
L. Fuhrmann and Dawei Xu for useful comments, 
and F. D'Ammando and J. Finke for providing their SED 
fit to PMN~J0948$+$0022. 
We also thank the anonymous referee for the helpful comments. 
This work is supported by NSFC grant no. 11473035, 11303046, 11273027 and by the Strategic Priority Research Program ``The Emergence of Cosmological Structures'' of the CAS (grant no. XDB09000000). 
This research has made use of the NASA/IPAC Infrared Science Archive and Extragalactic Database (NED), which are operated by the Jet Propulsion Laboratory, California Institute of Technology, under contract with the National Aeronautics and Space Administration. 
Funding for SDSS-III has been provided by the Alfred P. Sloan Foundation, the Participating Institutions, the National Science Foundation, and the U.S. Department of Energy Office of Science.



\bsp	
\label{lastpage}
\end{document}